\documentclass[amssymb,prb,twocolumn,showpacs]{revtex4}
\usepackage{epsfig}
\usepackage{dcolumn}
\usepackage{amsmath}
\begin{document}
\title{\bf Driving Weiss oscillations to Zero Resistance States by Microwave Radiation}
\author{Jes\'us I\~narrea$^{1,2}$ and Gloria Platero$^2$}
 \affiliation {$^1$Escuela Polit\'ecnica
Superior,Universidad Carlos III,Leganes,Madrid,Spain and  \\
$^2$Unidad Asociada al Instituto de Ciencia de Materiales, CSIC,
Cantoblanco,Madrid,28049,Spain.}
\date{\today}
%%%%%%%%%%%%%%%%%%%%%%%%%%%%%%%%%%%%%%%%%%%%%%%%%%%%%%%%%%%%%%%%%%
%%%%%%%%%%%%
\begin{abstract}
In this work we present a theoretical model to study the effect of
microwave radiation on Weiss oscillations. In our proposal Weiss
oscillations, produced by an spatial periodic potential, are
modulated by microwave radiation due to an interference effect
between both, space and time-dependent, potentials. The final
magnetoresistance depends mainly on the spatial period of the
spatial potential and the frequency of radiation. Depending on the
values of these parameters, we predict that Weiss oscillations can
reach zero resistance states. On the other hand, these
dissipationless transport states, created just by radiation, can be
destroyed by the additional presence of a periodic space-dependent
potential. Then by tuning the spatial period or the radiation
frequency, the magnetoresistance can be strongly modified.

\end{abstract}
%%%%%%%%%%%%%%%%%%%%%%%%%%%%%%%%%%%%%%%%%%%%%%%%%%%%%%%%%%%%%%%%%%
%%%%%%%%%%%%
\maketitle In the last two decades a lot of progress has been made
in the study of two-dimensional electron systems (2DES), and very
important and unusual properties have been discovered when these
systems are subjected to external potentials. Nowadays
magnetotransport properties of highly mobile 2DES are a subject of
increasing interest. In particular the study of the effects that
radiation can produce on these nano-devices is attracting
considerable attention both from theoretical and experimental
sides\cite{ina}. On the other hand, the interplay of two different
periodic modulations in a physical system will bring to interesting
features in dynamics and transport. For instance, the study of the
effect of microwave (MW) radiation on 2DES transport properties such
as Weiss oscillations \cite{weiss,ploog} represents a topic that
deserves to be studied\cite{dietel,torres}. Especially if we
consider that MW radiation gives rise also to magnetoresistance
($\rho_{xx}$) oscillations in 2DES and, at high MW intensity, zero
resistance states (ZRS)\cite{mani,zudov,studenikin,ina2}. Weiss
oscillations are also a type of $\rho_{xx}$ oscillations observed in
high mobility 2DES with a lateral periodic modulation (superlattice)
imposed in one direction. In this proposed scenario electrons are
subjected simultaneously to two different periodic potentials. One
is space-dependent, i.e., the superlattice, and the other is
time-dependent, i.e., MW radiation. Thus,  interesting properties
can be expected due to the combined effect of both potentials.
\begin{figure}
\centering\epsfxsize=3.5in \epsfysize=2.7in
\epsffile{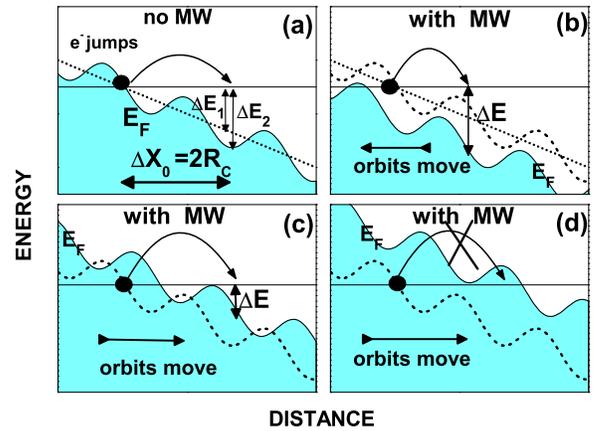} \caption{Schematic diagrams of electronic
transport of a 2DES with a spatial periodic modulation in one
direction. Thus, the Fermi energy presents a periodic and tilted
modulation in the transport direction. In Fig.(1.a) no MW field is
present. Compared to the case without superlattice ($\Delta E_{1}$),
more or less energy can be dissipated depending on the modulated
Fermi energy ($\Delta E_{2}$). When the MW field is on, the orbits
oscillate with $w$. In Fig.(1.b) the orbits move backwards during
the jump. In Fig.(1.c) the orbits are moving forwards but still the
electronic jump can take place. In Fig. (1.d) the orbits are moving
forwards but for a higher MW intensity, the combined effects of
spatial modulation and MW radiation make their amplitudes larger
than the electronic jump. Thus the electron movement between orbits
cannot take place because the final state is occupied. This
situation corresponds to ZRS.}
\end{figure}

In this letter we present firstly a theoretical approach to treat
Weiss oscillations. Secondly we introduce a more general model to
consider at the same footing the effects of a spatial periodic
modulation and MW radiation on the transport properties of a 2DES.
The former model is an alternative to the ones presented to date to
investigate Weiss oscillations. In these
models\cite{weiss,ploog,peeters} Weiss oscillations are explained in
terms of an oscillatory dependence of the bandwidth of the
modulation-broadened Landau levels of the 2DES. Beenakker presented
a model which considers a semiclassical approach\cite{bena}. In this
case the effect is explained as a resonance effect between the orbit
motion and an oscillating drift produced by magnetic field and the
electric field imposed by the unidirectional periodic potential. In
the proposal presented here the origin of Weiss oscillations is a
periodically modulated Fermi energy as a direct consequence of the
spatial periodic potential $V(x)$ imposed on the 2DES. The spatially
modulated Fermi energy changes dramatically the scattering
conditions between electrons and charged impurities presented in the
sample. Depending on the external DC magnetic field ($B$), compared
to the case without superlattice modulation, at certain $B$ the
transport will be more dissipative. This corresponds to a peak in
$\rho_{xx}$. Meanwhile in others, the transport will be less
dissipative corresponding to a $\rho_{xx}$ valley. Under this
scenario it is expected that the presence of MW radiation\cite{ina3}
will alter dramatically the $\rho_{xx}$ response of the system. We
predict that the combined effect of MW radiation and spatial
potential  will lead the system to a interference regime with
constructive and destructive responses. As a consequence $\rho_{xx}$
will present a modulated profile and {\it for a sort of external
parameters, spatial modulation and MW frequency ($w$), ZRS can be
created or destroyed}. The theory presented here can have potential
applications in other fields and systems like nano
electro-mechanical systems (NEMS)\cite{pisto} with AC-potentials and
surfaces acoustic waves (SAW)\cite{saw} in 2DES or dots illuminated
with MW radiation. In other words, the physics presented in this
letter can be of interest, from a basic physics standpoint, for an
audience dealing with the effects that AC or/and DC fields produce
on nano-devices.
\begin{figure}
\centering\epsfxsize=3.5in \epsfysize=3.0in
\epsffile{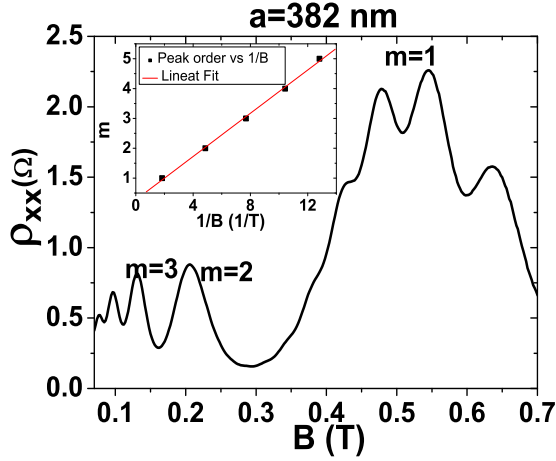} \caption{Calculated $\rho_{xx}$ as a
function of magnetic field $B$ ({\it Weiss oscillations}). The
period of the static modulation is $a=382 nm$ and the modulation
amplitude is $V_{0}\sim 0.1 meV$. The inset shows the linear
dependence of the peak index $m$ vs $1/B$ which means that Weiss
oscillations are periodic in $1/B$. T=1K.}
\end{figure}

Our system consists in a 2DES subjected to a perpendicular $B$
(z-direction) and a DC electric field ($E_{dc}$) (x-direction). We
include the unidireccional periodic potential $V(x)=V_{0} \cos Kx$
where $K=2\pi/a$, $a$ being the spatial period of the superlattice.
The total hamiltonian $H$, can be written as:
\begin{eqnarray}
H&=&\frac{P_{x}^{2}}{2m^{*}}+\frac{1}{2}m^{*}w_{c}^{2}(x-X_{0})^{2}-eE_{dc}X_{0}+\frac{1}{2}m^{*}\frac{E_{dc}^{2}}{B^{2}}\nonumber \\
 & &+V_{0} \cos (Kx) =H_{0}+V_{0} \cos (Kx)
\end{eqnarray}
$X_{0}$ is the center of the orbit for the electron spiral motion:
%\begin{equation}
$X_{0}=\frac{\hbar k_{y}}{eB}- \frac{eE_{dc}}{m^{*}w_{c}^{2}}$,
%\end{equation}
$e$ is the electron charge, $w_{c}$ is the cyclotron frequency.
$H_{0}$ is the hamiltonian of a harmonic quantum oscillator and its
wave functions, the well-known oscillator functions (hermite
polynomials). We treat $V_{0} \cos (Kx)$ in first order perturbation
theory and the first order energy correction is given by:
$\epsilon_{n}^{(1)}= V_{0} \cos (KX_{0})e^{-X/2}L_{n}(X)=U_{n}\cos
(KX_{0})$, where $X=\frac{1}{2}l^{2}K^{2}$, $L_{n}(X)$ is a Laguerre
polynomial and $l$ the characteristic magnetic length. Therefore the
total energy for the Landau level $n$ is given by:
$\epsilon_{n}=\hbar w_{c}(n+\frac{1}{2})-eE_{dc}X_{0}+U_{n}\cos
(KX_{0})$. This result affects dramatically the Fermi energy as a
function of distance (see Fig. 1), showing now a periodic, tilted
modulation.

Now we introduce impurity scattering suffered by the electrons in
our model \cite{ina2,ina3,ridley}. When one electron scatters
elastically due to charged impurities, its average orbit center
position changes, in the electric field direction, from $X_{0}$ to
$X_{0}^{'}$. Accordingly the average advanced distance in the $x$
direction is given by $\Delta X_{0}=X_{0}^{'}-X_{0}\simeq 2R_{c}$,
$R_{c}=\frac{\sqrt{2m^{*}E_{F}}}{eB}$ being the orbit radius (see
Fig. 1). Without the static periodic modulation $V(x)$, this
distance corresponds to an energy increase regarding the Fermi
energy given by $\Delta \epsilon=\epsilon_{n}-\epsilon_{n^{'}}\simeq
eE_{dc}2R_{c}$. This energy is eventually dissipated by the lattice
being responsible of the magnetoresistance measured in the sample.
Thus, there is a direct relation between advanced distance and
dissipated energy. However if the 2DES is subjected to $V(x)$, the
energy increase has a different expression:
\begin{equation}
\Delta \epsilon\simeq eE_{dc}2R_{c}+U_{n}[\cos (KX_{0})-\cos
(KX_{0}^{'})]
\end{equation}
To evaluate the {\it average energy change} in the scattering jump,
we consider that for the initial position, $X_{0}$, the electron has
the same energy as without $V(x)$ (see Fig 1a). Thus, the net energy
change in the jump comes only from the final position $X_{0}^{'}$.
This implies that $\cos KX_{0}=0$ $\Rightarrow$ $KX_{0}=(2m+1)\pi/2$
with $m=0,1,2,3...$. Taking for simplicity the case $m=0$,
$\Rightarrow KX_{0}^{'}=2R_{c}K+\pi/2$, and considering the
asymptotic relation for $L_{n}(X)$\cite{grad} (for large values of
$n$), $L_{n}(X)\rightarrow J_{0}(2\sqrt{X})$ we can finally write
for $\Delta \epsilon$:
\begin{equation}
\Delta \epsilon\simeq eE_{dc}2R_{c}- V_{0}e^{-X/2} J_{0}(2\sqrt{X})
\cos\left[ 2\left(R_{c}K+\frac{\pi}{4}\right)\right]
\end{equation}
being $J_{0}$ Bessel function of zero order. According to this
expression the energy difference regarding the Fermi energy is now
depending of $\frac{1}{B}$ through a cosine function. Thus, it can
be larger or smaller than the situation without static modulation
(see Fig. 1a). A larger or smaller energy is now dissipated giving
rise to an increasing or decreasing $\rho_{xx}$,i.e.,  $\rho_{xx}$
oscillations. The advanced distance corresponding to $\Delta
\epsilon$ can be calculated straightforward:
\begin{figure}
\centering\epsfxsize=3.0in \epsfysize=3.0in
\epsffile{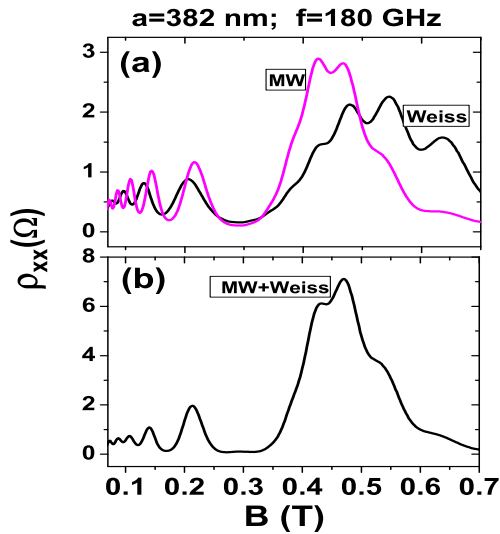} \caption{Calculated $\rho_{xx}$ vs $B$. In
Fig. 3a, the Weiss-labeled curve corresponds to the 2DES with an
spatial periodic potential (same parameters as Fig. 2) imposed on
top of it, but not illuminated with MW. The MW-labeled curve
corresponds to the same 2DES without spatial periodic potential but
being illuminated with MW. None of them present ZRS. In Fig. 3b, the
sample is subjected to both potentials and ZRS are obtained around
$B=0.3T$. (T=1K.)}
\end{figure}
\begin{figure}
\centering\epsfxsize=3.0in \epsfysize=3.5in
\epsffile{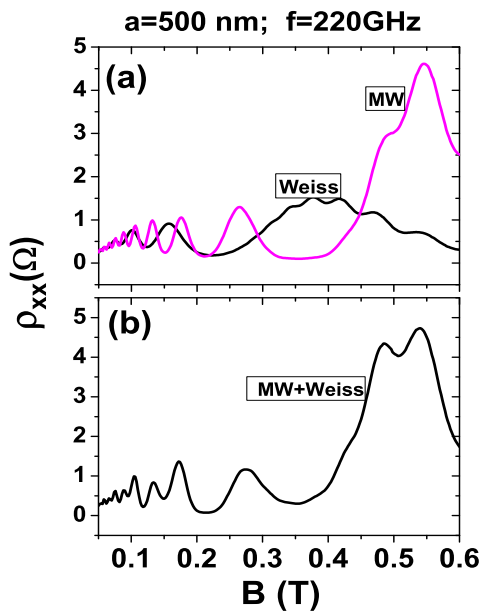} \caption{Calculated $\rho_{xx}$ vs $B$.
Similar panels as in Fig. 3. We observe creation ($B=0.2T$) and
destruction ($B=0.35T$) of ZRS due to the combined effect of both
periodic potentials. (T=1K).}
\end{figure}
\begin{equation}
\Delta X_{T}=2R_{c}- \frac{V_{0}e^{-X/2} J_{0}(2\sqrt{X})}{eE_{dc}}
\cos\left[ 2\left(R_{c}K+\frac{\pi}{4}\right)\right]
\end{equation}
\begin{figure}
\centering\epsfxsize=3.5in \epsfysize=5.0in
\epsffile{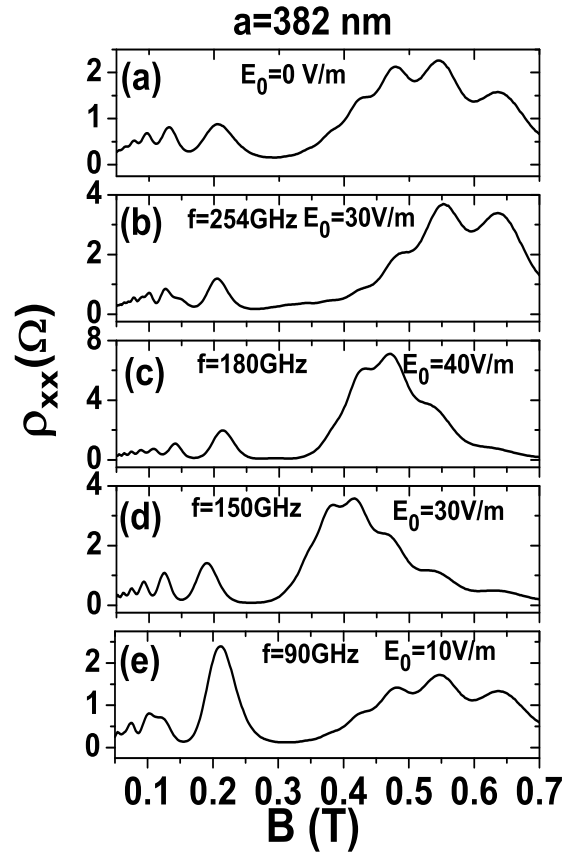} \caption{Calculated $\rho_{xx}$ vs $B$ for
a 2DES with a spatial periodic potential (same parameters as in Fig.
2) under different regimes. In Fig. 5a there is no MW field. From
Fig. 5b to 5e we present calculated results with the MW field on and
for four different frequencies. (T=1K.).}
\end{figure}
Following references [9,12,15] from $\Delta X_{T}$ we calculate a
drift velocity for the electron dissipative transport in the $x$
direction and finally we obtain $\rho_{xx}$. According to this
expression the minima values of $\rho_{xx}$ are obtained for
$2R_{c}=[m-\frac{1}{4}]a$, (commensurability condition)\cite{weiss}.
The maxima for $2R_{c}=[m+\frac{1}{4}]a$. In Fig. 2 we represent
calculated $\rho_{xx}$ as a function of $B$. The period of the
static modulation is $a=382$ nm and the modulation amplitude is
$V_{0}\sim 0.1 meV$. The substructure appearing at higher $B$
corresponds to the Shubnikov-deHaas oscillations. We reproduce Weiss
oscillations with reasonable agreement with
experiments\cite{weiss,ploog}.

If now we switch on the MW field, we expect that its effect will
alter dramatically the $\rho_{xx}$ response of the system. From the
MW driven Larmor orbits model\cite{ina2,ina3} radiation forces the
electronic orbits to move back and forth at the frequency of the
field. This affects the scattering conditions of the 2DES increasing
or decreasing the distance of the scattering jump giving rise to
MW-induced $\rho_{xx}$ oscillations. Both effects, periodic spatial
modulation and MW radiation alter simultaneously the average
advanced distance when the electron scatters (see Fig. 1). We obtain
an expression for the total average distance in the $x$ direction:
\begin{equation}
\Delta X_{T}=2R_{c}- S\cos\left[
2\left(R_{c}K+\frac{\pi}{4}\right)\right]+A \cos (w \tau)
\end{equation}
where
%\begin{equation}
$S=\frac{V_{0}e^{-X/2} J_{0}(2\sqrt{X})}{eE_{dc}}$
%\end{equation}
and $A=\frac{e
E_{o}}{m^{*}\sqrt{(w_{c}^{2}-w^{2})^{2}+\gamma^{4}}}$. $E_{o}$ is
the MW electric field amplitude, $\gamma$ is a phenomenologically
introduced\cite{ina2} damping parameter and $\tau$ the scattering
time. As in the previous case without MW, from $\Delta X_{T}$ we
calculate an electron drift velocity\cite{ina2,ina3,ridley} and
finally $\rho_{xx}$ of the system. According to our model there is a
direct relationship between $\Delta X_{T}$ and $\rho_{xx}$:
\begin{equation}
\rho_{xx} \propto -S\cos\left[
2\left(R_{c}K+\frac{\pi}{4}\right)\right]+A \cos (w \tau)
\end{equation}
Therefore from equation 6 we can predict an interference regime
between both periodic potentials which will be reflected on
$\rho_{xx}$\cite{ina4}. That means that depending mainly on $a$, $w$
and for some values of $B$, $\rho_{xx}$ will present a constructive
response and a reinforced signal. Accordingly ZRS can be found where
they did not exist before. Also we can have in some points much
higher current than the dark case. For other values of $B$ the
interference will be destructive and the $\rho_{xx}$ response will
be closer to the one of the dark case. Also the destructive
interference can get rid of ZRS that MW radiation can previously
create. In other words, ZRS created by the action of MW radiation on
a 2DES can be destroyed by the additional presence of a spatial
periodic potential on the system.
%Thus, we
%predict as well that for a certain values of $a$ and $w$ and due to
%the combined effect of both potentials, ZRS. Although for only one
%potential, ZRS did not show up before.

In Fig. 3a we present calculated $\rho_{xx}$ vs $B$ for two
different regimes of the same 2DES. In one (Weiss-labeled) the
system has a superlattice (parameters of Fig. 2) imposed in the
transport directions but it is not illuminated with MW. We obtain
the well-known Weiss oscillations. In the other (MW-labeled) the
system does not present the superlattice but it is illuminated with
MW. We obtain the also well-known MW-induced resistance
oscillations. In none of them ZRS show up. In Fig. 3b we present
calculated $\rho_{xx}$ vs $B$ showing the combined effect of both
potentials. We can see clearly that ZRS show up around $B=0.3 T$. In
Fig. 4  we present similar panels as in Fig. 3 but for different $a$
and $w$. We observe in the bottom panel two opposite effects in
terms of ZRS: the combined effect of both periodic potentials are
able to create ZRS where they did not exist before (around 0.2 T) or
destroy them where they existed (around 0.35 T). In Fig. 5 we can
see calculated $\rho_{xx}$ vs $B$ for a 2DES with a spatial periodic
potential of $a=382$ nm, under different regimes. In Fig. 5a we
present same results as in Fig. 2 just for comparison. From Fig. 5b
to 5e we present calculated results with the MW field on and for
four different frequencies. Comparing the bottom panels to the top
one we can see the different interference effects.

In summary, we have analyzed the interplay between a spatially
modulated periodic potential and a time-dependent periodic field and
its effect on the magnetotransport in a 2DES. In particular, we have
presented a theoretical approach to investigate the effect of
microwave radiation on Weiss oscillations. In our proposal Weiss
oscillations are modulated by microwave radiation due to an
interference effect between both, spatial and time-dependent,
periodic potentials. The final profile presented by $\rho_{xx}$
depends mainly on the spatial period of the superlattice modulation
and the frequency of MW.

This work has been supported by the MCYT (Spain) under grant
MAT2005-0644, by the Ram\'on y Cajal program and  by the EU Human
Potential Programme: HPRN-CT-2000-00144.

\end{document}